\documentclass[
 aip,
 amsmath,amssymb,
 reprint,%
]{revtex4-1}

\usepackage{ragged2e} 
\usepackage{subfig}
\usepackage{graphicx}
\usepackage{dcolumn}
\usepackage{bm}
\usepackage[utf8]{inputenc}
\usepackage[T1]{fontenc}
\usepackage{mathptmx}
\usepackage{etoolbox}

\makeatletter
\def\@email#1#2{%
 \endgroup
 \patchcmd{\titleblock@produce}
  {\frontmatter@RRAPformat}
  {\frontmatter@RRAPformat{\produce@RRAP{*#1\href{mailto:#2}{#2}}}\frontmatter@RRAPformat}
  {}{}
}%
\makeatother

\begin{document}
\preprint{AIP/123-QED}

\title[Manarat: A Scalable QICK-Based Control System for Superconducting Quantum Processors]{Manarat: A Scalable QICK-Based Control System for Superconducting Quantum Processors \\ Supporting Synchronized Control of 10 Flux-Tunable Qubits}

\author{Agustin Silva}
 \email{agustin.silva@tii.ae}
\author{Alvaro Orgaz-Fuertes}
 \email{alvaro.orgaz@tii.ae}
 
\affiliation{Quantum Research Center, Technology Innovation Institute, Abu Dhabi, UAE}

\date{\today}

\begin{abstract}
A scalable control architecture for superconducting quantum processors is essential as the number of qubits increases and coherent multi-qubit operations span beyond the capacity of a single control module. The Quantum Instrumentation Control Kit (QICK), built on AMD RFSoC platforms, offers a flexible open-source framework for pulse-level qubit control but lacks native support for multi-board synchronization, limiting its applicability to mid- and large-scale quantum devices. To overcome this limitation, we introduce Manarat, a scalable multi-board control platform based on QICK that incorporates hardware, firmware, and software enhancements to enable sub-100 ps timing alignment across multiple AMD ZCU216 RFSoC boards. Our system integrates a low-jitter clock distribution network, modifications to the tProcessor, and a synchronization scheme to ensure deterministic alignment of program execution across boards. It also includes a custom analog front-end for flux control that combines high-speed RF signals with software-programmable DC biasing voltages generated by a low-noise, high-precision DAC. These capabilities are complemented by a software stack capable of orchestrating synchronized multi-board experiments and fully integrated with the open-source Qibo framework for quantum device calibration and algorithm execution. We validate Manarat on a 10-qubit superconducting processor controlled by two RFSoC boards, demonstrating reliable execution of synchronized control sequences for cross-board CZ gate calibration. These results confirm that sub-nanosecond synchronization and coherent control is achievable across multiple RFSoC boards, enabling scalable operation of superconducting quantum computers. 
\end{abstract}

\maketitle

\section{Introduction}
Quantum computers promise to solve certain problems exponentially faster than classical systems \cite{shor1999polynomial, arute2019quantum}, but unlocking this potential requires scaling up the number of qubits while maintaining precise, high-fidelity control \cite{preskill2018quantum, kjaergaard2020superconducting}. Superconducting qubits are among the leading platforms \cite{devoret2013superconducting}, yet they face significant challenges from noise, decoherence, and crosstalk \cite{gambetta2017building}. One approach to mitigate these issues is the use of quantum error correction, which encodes a single high-performance logical qubit using many noisy physical qubits \cite{gottesman1997stabilizer, fowler2012surface, terhal2015quantum}.

Another strategy to improve quantum processor performance involves the use of tunable couplers, which help minimize unwanted interactions and reduce crosstalk between qubits \cite{chen2014qubit, mundada2019suppression, sung2021realization}. However, both strategies significantly increase the number of control channels required \cite{versluis2017scalable}. This intensifies the demand for scalable and precise control electronics, which have become the most expensive component in a superconducting qubit quantum computing laboratory \cite{krinner2019engineering, mcdermott2014accurate}.

In recent years, open-source control platforms have emerged as a promising strategy to reduce the cost and complexity of quantum control systems \cite{xu2021qubic, xu2023qubic}. One of the most prominent examples is the Quantum Instrumentation Control Kit (QICK), developed at Fermilab \cite{stefanazzi2022qick, ding2024experimental}. QICK is built around AMD RFSoC (Radio Frequency System-on-Chip) devices \cite{amd_rfsoc_overview}, which integrate high-speed data converters, programmable logic, and ARM embedded processors on a single chip. The platform supports direct digital synthesis of microwave pulses, eliminating the need for frequency upconversion via local oscillators and IQ mixers, and significantly simplifying the control stack \cite{werninghaus2021leakage, gustavsson2013improving}.

QICK has enabled a range of high-impact experiments across superconducting qubit platforms. Bland et al.\cite{bland20252d} demonstrated transmon qubits with energy-relaxation and dephasing times ($T_1$ and $T_2$) exceeding 1 ms and single qubit gate fidelities above 99.99\%, representing a major improvement in coherence and control. Furthermore, Wang et al. \cite{wang2025evidence} reported the first Floquet-mode traveling-wave parametric amplifier fabricated within a superconducting-qubit process, achieving >20 dB gain over a 3 GHz bandwidth, with a quantum efficiency of 92 ± 7\% and record system measurement efficiency of 65 ± 6\%.

Despite its flexibility and performance, QICK is currently limited by the absence of native multi-board synchronization capabilities. While boards such as AMD ZCU216 \cite{amd_zcu216}, which have 16 channels of 14-bit, 2.5 GSPS ADCs, and 16 channels of 14-bit, 9.85 GSPS DACs, provide a considerable amount of input and output channels, they are insufficient for midscale quantum processors. For example, controlling a 10-qubit flux-tunable transmon device in a ladder configuration requires 10 drive lines, 10 flux lines, and 2 readout lines, which exceed the resources of a single RFSoC board.

To overcome this limitation, we introduce \textbf{Manarat}, a scalable control platform that builds on the QICK architecture with notable modifications to hardware, firmware, and control software. This control platform enables sub-100 ps synchronization across multiple RFSoC boards, allowing coherent multi-qubit control with a distributed system architecture.

Achieving this level of synchronization across FPGA boards using the QICK framework requires overcoming three key challenges. First, a low-jitter reference clock must be distributed to all participating boards to establish a shared timing baseline. Second, a synchronization instruction must align each independent board's program, ensuring precise timing at the onset of every pulse sequence. Third, each board must detect and respond to the synchronization trigger in the same clock cycle, even when executing different programs.

In this work, we describe the design and implementation of our extended platform and present experimental benchmarks that demonstrate its timing alignment and suitability for operating superconducting quantum processors beyond the single-board limit.

\section{Related Work}
Recent work by Xu et al. \cite{xu2025multi}, developed as part of the QubiC project at Berkeley Lab, introduced a multi-board synchronization and communication framework for RFSoC-based quantum control. Their system achieves deterministic clock alignment and low-latency data transfer across multiple RFSoC boards using a ring-based synchronization protocol and fiber optic links, supporting mid-circuit measurement and feedforward logic. The architecture was validated using a room temperature qubit readout emulator.

In contrast, our work focuses on addressing just the fundamental challenge of scalable, synchronized control across multiple RFSoC boards but is specifically designed and experimentally validated for superconducting qubit processors. It achieves sub-100 ps timing alignment between RFSoC boards and supports synchronized microwave pulse generation for multi-qubit calibration and two-qubit gate operations on a 10-qubit processor. In addition, our control stack incorporates a custom analog front-end for flux pulses, and includes firmware and software features tailored to the fast, high-fidelity requirements of superconducting quantum hardware.

\section{System Architecture and Enhancements} \label{sec:System_Design}

This section presents an overview of QICK architecture as a baseline platform for superconducting qubit control, followed by a detailed description of the hardware, firmware, and software modifications implemented to scale the control system beyond a single board.

\subsection{QICK Overview}

The QICK platform provides a reference architecture for implementing qubit control and readout leveraging the high-performance resources of AMD’s RFSoC devices. Its control logic is built primarily using three core firmware components: signal generators for pulse synthesis, readouts for signal acquisition and processing, and a custom timing processor (tProcessor) for scheduling the operations of the signal generators and readouts.

QICK exposes a Python software interface for the definition and execution of experiments. QICK is built on top of PYNQ \cite{pynq_website}, an open-source project from AMD that enables developers to program Zynq System-on-Chip (SoC) devices and interface with their firmware using Python instead of traditional embedded C or C++ code. The Python interpreter runs on a Linux distribution operating on one of the ARM processors embedded within the RFSoC.

QICK includes three types of configurable \textbf{signal generators} connected to the RF DACs: 

\begin{itemize}
    \item \textit{Full-speed generators} operate at the maximum sampling rate of the DAC (up to 9.85 GS/s on the ZCU216), using digital carriers through direct digital synthesis (DDS) modulated by arbitrary waveform envelopes. These are ideal for high-fidelity gates requiring precise pulse shaping with sub-nanosecond resolution. 
    \item \textit{Interpolated generators} reduce FPGA resource usage by generating the envelope at a lower sample rate (1/16 of the DAC rate) and interpolating the signal digitally. 
    \item \textit{Multiplexed generators} digitally combine multiple DDS channels before the DAC output, enabling simultaneous multitone control for frequency multiplexed qubits using a single output channel.
\end{itemize}

On the acquisition side, QICK processes the signals acquired by the RF ADCs using two types of \textbf{readout blocks}:

\begin{itemize}
    \item \textit{standard readout} blocks perform a digital downconversion (DDC) to bring the signal to the baseband, followed by a low-pass filter and a decimation by a factor of 8. The results of multiple acquisitions can be integrated and averaged in a buffer. These blocks are optimized for single tone and long integration times.
    \item \textit{multiplexed readout} blocks implement a polyphase filter bank (PFB) to demultiplex the digitized signal into multiple subbands, each with its own DDS/NCO for tone-specific demodulation. This enables simultaneous readout of multiple qubits using a single ADC stream, with support for up to 8 channels per stream.
\end{itemize}

Sequencing is managed by the timing processor (tProcessor), a lightweight custom CPU implemented in the FPGA fabric that executes low-level custom assembly instructions to schedule pulses and acquisitions, manipulate registers, implement loops and conditional branching. The instructions are executed in a fixed number of cycles, enabling deterministic and cycle-accurate control. The signal generators are triggered by the tProcessor, and the waveform parameters are dynamically updated via register writes.

The tProcessor supports execution of loops and dynamic parameter updates directly within the FPGA, enabling what is referred to as real-time execution. In this mode, the entire experiment is managed within the FPGA, with minimal latency and overhead, as no external communication is required during the experiment's execution. Although some parameters cannot be dynamically modified by the tProcessor and instead require loop control from Python running on the ARM processor, the tight integration between the processing system and the programmable logic of the RFSoC chip ensures that the communication overhead between software and hardware remains exceptionally low. This mode of operation is often referred to as near-real-time execution.

QICK provides PYNQ drivers for all of these IP blocks so that they can be configured directly in Python. The user interface layer includes utilities for defining pulse shapes, scheduling pulse sequences, and reading out results. Control programs written in Python are compiled into a custom assembly language for the tProcessor and uploaded along with waveform data and configuration registers.

While QICK is well-suited for single-board operation, its current architecture assumes local execution and coordination, with each board's tProcessor running independently. Although the design includes preliminary mechanisms such as external triggering, there is no native support for global clock distribution, processor synchronization or distribution of programs across multiple boards. Features such as SYSREF alignment, MTS configuration, and multi-board experiment orchestration are left to the implementer. 

The XM655 board \cite{amd_xm655}, the analog front-end included with the ZCU216 evaluation board, is designed for telecommunication applications and is not suitable for controlling superconducting qubits. To address this problem, Fermilab has developed a custom analog front-end called QICK RF board. In addition to adapting the raw signals from data converters, it includes tunable low-noise amplification and tunable filtering, as well as high-precision DACs for biasing flux-tunable qubits. Unfortunately, we were not able to procure those boards in time for the experiments presented in this paper, and we implemented a simplified custom front-end suitable for precise flux control, based on available components.

In summary, QICK provides a flexible and accessible baseline architecture for qubit control on RFSoC platforms that combines low-level extensible firmware primitives with a Python software interface. Our work builds on this architecture by addressing its key limitations in synchronization, enabling multi-board scalable operation suitable for mid-scale superconducting qubit systems.

\subsection{Hardware}

\begin{figure*}
    \centering
    \includegraphics[width=0.8\linewidth]{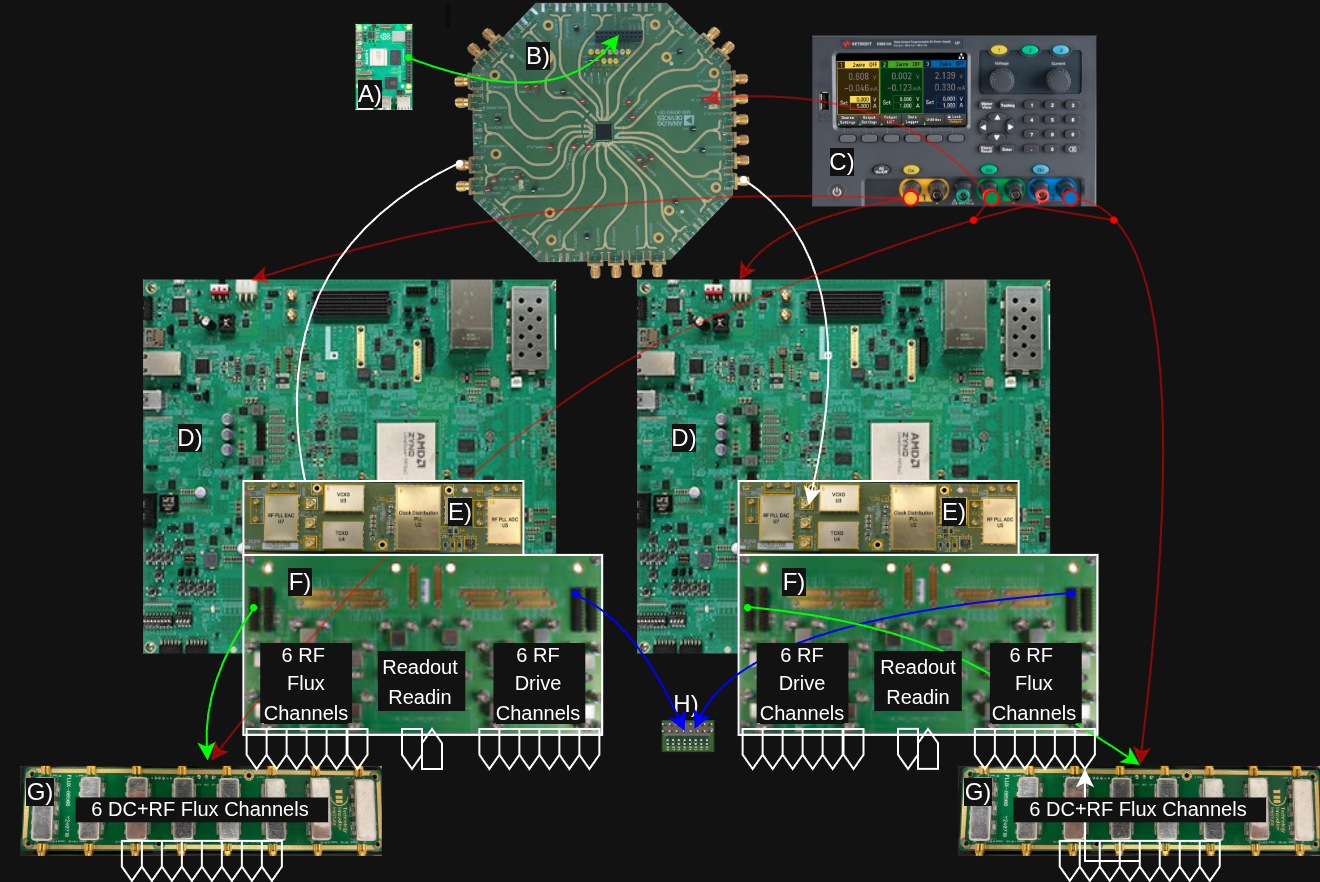}
    \caption{Setup for synchronizing two boards to control up to 12 flux-tunable qubits. A) Rasberry Pi. B) Analog Devices HMC7044 clock generator. C) Keysight E36313A power supply. D) AMD ZCU216 RFSoC evaluation board. E) CLK104 containing LMK04828 clock generator. F) XM655 analog front-end. G) Custom flux analog front-end. H) Custom synchronization signaling bus. Color-coded connections between devices: red = POWER, white = SYSREF, green = SPI configuration, blue = SYNC signal.}
    \label{fig:setup.drawio}
\end{figure*}

Although our design theoretically supports synchronization of up to six boards without additional hardware, we have so far validated the approach using a two-board configuration, which is used to control 10 flux-tunable superconducting qubits. The hardware setup is illustrated in Figure \ref{fig:setup.drawio} and includes the following components:

\begin{itemize}
    \item 2x AMD ZCU216 evaluation boards with their corresponding CLK104 and XM655 boards
    \item 2x custom flux analog front-end boards
    \item 1x custom synchronization signaling bus board
    \item 1x Analog Devices HMC7044 evaluation board
    \item 1x Raspberry Pi V
    \item 1x Keysight E36313A power supply
\end{itemize}

This configuration enables the control of:

\begin{itemize}
    \item 12 RF drive lines
    \item 12 combined DC+RF flux lines
    \item 2 readout lines (each supporting up to 8 frequency-multiplexed tones)
    \item 2 readin lines (each supporting up to 8 frequency-multiplexed tones)
\end{itemize}

\subsubsection{XM655 Analog Front-End}

To condition RF signals for qubit drive and readout, we use the XM655 front-end module bundled with the ZCU216 evaluation board. XM655 add-on cards include baluns at various frequencies and SMA connectors for interfacing with DACs and ADCs. This front-end includes a limited set of baluns that span different frequency ranges: four optimized for 4–5 GHz and four for 5–6 GHz. The qubits in our system have first transition frequencies in the 4–5 GHz range, with approximately half centered near 4.2 GHz and the remainder near 4.8 GHz. We allocate all four 4–5 GHz baluns for qubit drive lines and assign one 5–6 GHz balun on each board to one of the high frequency qubits.

Although some qubit frequencies lie outside the nominal bandwidth of the assigned baluns, we observe no significant limitations in control performance. While off-band attenuation is increased, we verify that the available signal power remains sufficient for effective qubit control. Similarly, the readout resonators operate around 7.5 GHz, beyond the upper limit of the 5–6 GHz baluns used for their drive lines. Despite this mismatch, we achieved a reliable readout with adequate signal-to-noise ratio.

RF signals for flux control are routed from the positive (P) outputs of the ZCU216 DACs directly into our custom analog front-end. This interface removes the DC component of the signal using integrated bias tees and delivers fast AC pulses superimposed on DC bias voltages. 

\subsubsection{HMC7044 Clock Generator}

The Analog Devices HMC7044 \cite{analog_devices_hmc7044} is a dual-loop jitter attenuating clock generator and distributor designed to deliver low phase noise and deterministic clocking for high-speed data converters. In our setup, we employ its evaluation board, which is readily available and provides up to six clock outputs, although the chip itself supports a total of 14 configurable outputs.

To achieve deterministic phase alignment and synchronization across multiple RFSoC-based boards, we configure the HMC7044 to generate and distribute a low-frequency reference clock at 7.68 MHz to each board. Each FPGA subsequently utilizes this reference clock to generate its necessary higher-frequency internal clocks, ensuring coherent operation across the system.

\begin{figure}[!ht]
    \centering
    \includegraphics[width=0.9\linewidth]{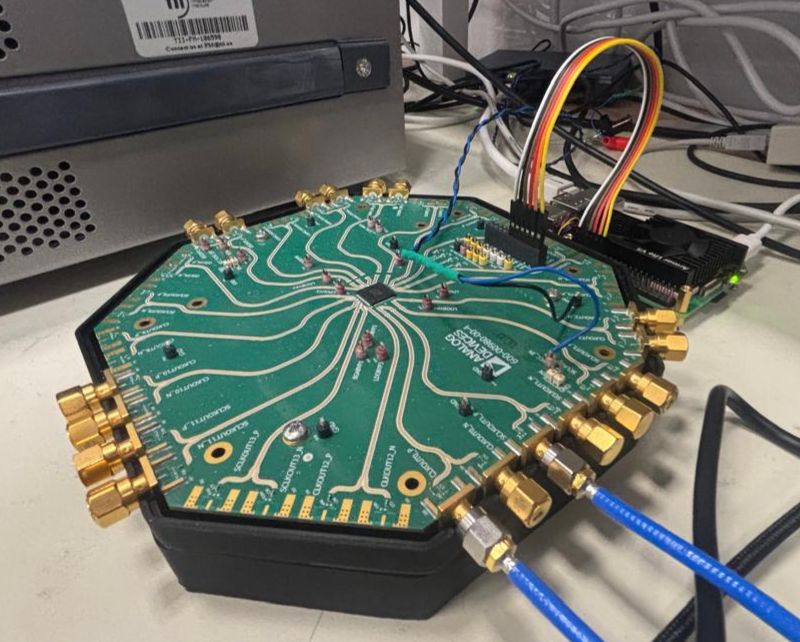}
    \caption{Analog Devices HMC7044 clock distribution module (central board in the photo) together with a Raspberry Pi (upper-right side of the photo) for programming and monitoring purposes.}
    \label{fig:HMC7044}
\end{figure}

A Raspberry Pi is used to automate the configuration and monitoring of the clock distribution system, Fig. \ref{fig:HMC7044}. It serves two main functions:

\begin{itemize}
    \item \textit{Startup Configuration:} The Raspberry Pi initializes the HMC7044 via its SPI interface, using its onboard GPIO pins. This configuration is executed at system startup to ensure the proper distribution of the reference clock.
    \item \textit{Synchronization Monitoring:} The Raspberry Pi continuously monitors the synchronization status by connecting to all RFSoC boards over SSH. It verifies that each board’s PLL remains locked to the reference outputs of the HMC7044. If a loss of lock is detected, the Raspberry Pi can automatically reprogram the HMC7044 to restore proper synchronization.
\end{itemize}

\subsubsection{LMK04828 Clock Generator}

\begin{figure}[!ht]
    \centering
    \includegraphics[width=\linewidth]{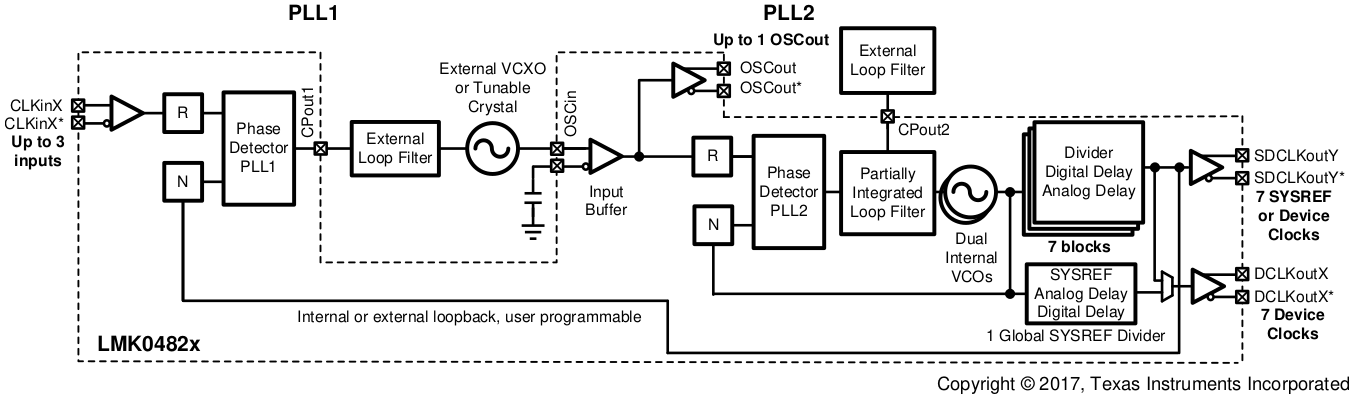}
    \caption{Simplified functional block diagram illustrating the nested 0-delay dual loop mode configuration of the LMK04828.}
    \label{fig:LMK04828}
\end{figure}

The CLK104 boards included with the ZCU216 platform host multiple programmable clock generation and conditioning devices, most notably the LMK04828 \cite{ti_lmk04828}, a high-performance clock conditioner supporting a range of operating modes. In our setup, we configure it to be in nested zero-delay dual-loop mode, Fig. \ref{fig:LMK04828}.

This mode ensures that all generated clock outputs maintain a deterministic phase relationship relative to the input reference signal provided by the HMC7044, allowing for tight synchronization across multiple boards.

The LMK04828 is configured over I2C from the RFSoC's processing system using thorugh the \textit{xrfclk} \cite{xilinx_pynq_xrfclk} Python library. It receives the 7.68 MHz reference clock generated by the HMC7044 and produces the following synchronized output clocks:

\begin{itemize}
    \item DCLKout6 and DCLKout12: 245.76 MHz outputs for DAC and ADC PLL inputs, respectively, used to generate high-speed converter sampling rates of 5.89824 GSPS for DACs and 2.4576 GSPS for ADCs.
    \item DCLKout8: 122.88 MHz primary FPGA design clock, fully synchronized across all boards, used for AXI and QICK IP clock domains.
    \item SCLKout9: 7.68 MHz SYSREF signal, required for Multi-Tile Synchronization (MTS).
\end{itemize}

\subsubsection{Custom Flux Analog Front-End}

\begin{figure}[!ht]
    \centering
    \includegraphics[width=\linewidth]{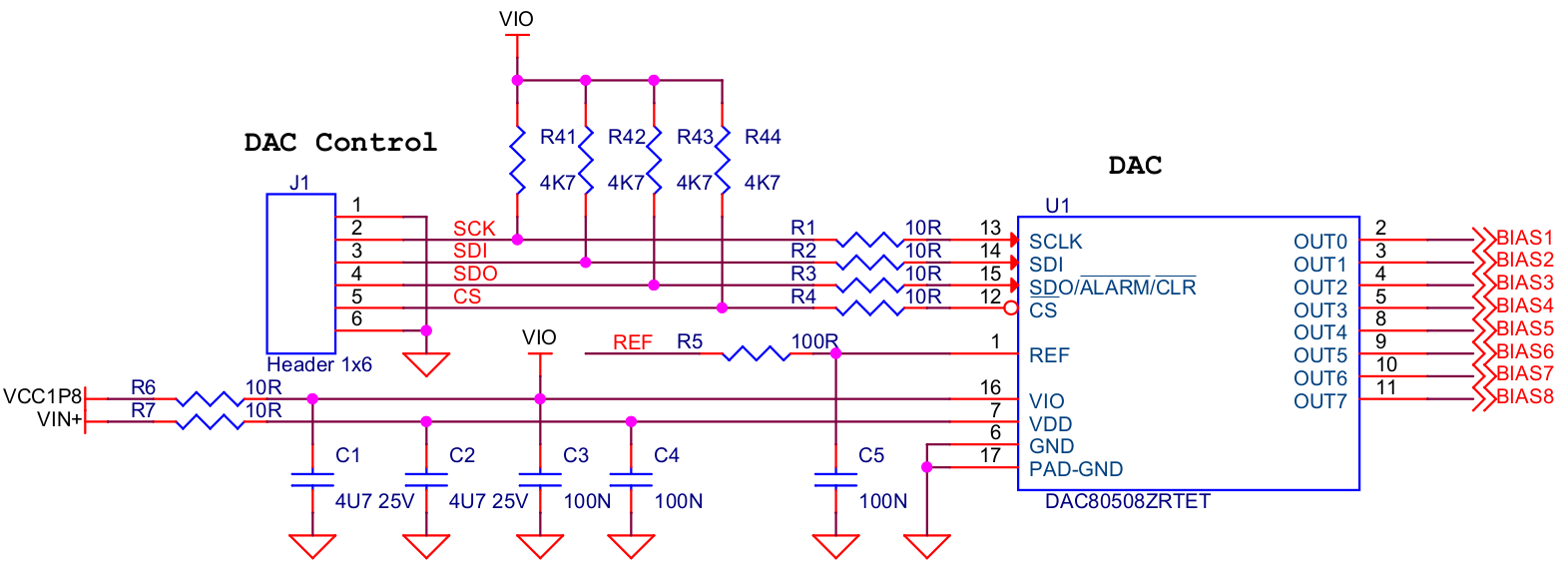}
    \caption{Schematic for the slow DC DAC control.}
    \label{fig:DC_flux_1}
\end{figure}

We developed a custom analog front-end (AFE) to generate and combine direct current (DC) and radio frequency (RF) signals required for the control of flux-tunable superconducting qubits and couplers. Flux control requires stable, high-resolution DC biasing to establish and maintain qubit operating frequencies over long timescales. The standard analog front-end of the ZCU216 platform is not suitable for this application, as it provides only AC-coupled outputs with a limited voltage range. To address these limitations, we integrated a Texas Instruments DAC80508, see Fig. \ref{fig:DC_flux_1}, a low-noise 16 bit digital-to-analog converter with eight independent output channels, each capable of supplying adjustable voltages up to ±2.5 V. This enables precise flux control across a wide frequency range.

DC bias voltages are dynamically programmed through our Python-based SPI driver interface running on the ZCU216 processing system (PS) to facilitate straightforward integration into the control software stack.

\begin{figure}[!ht]
    \centering
    \includegraphics[width=\linewidth]{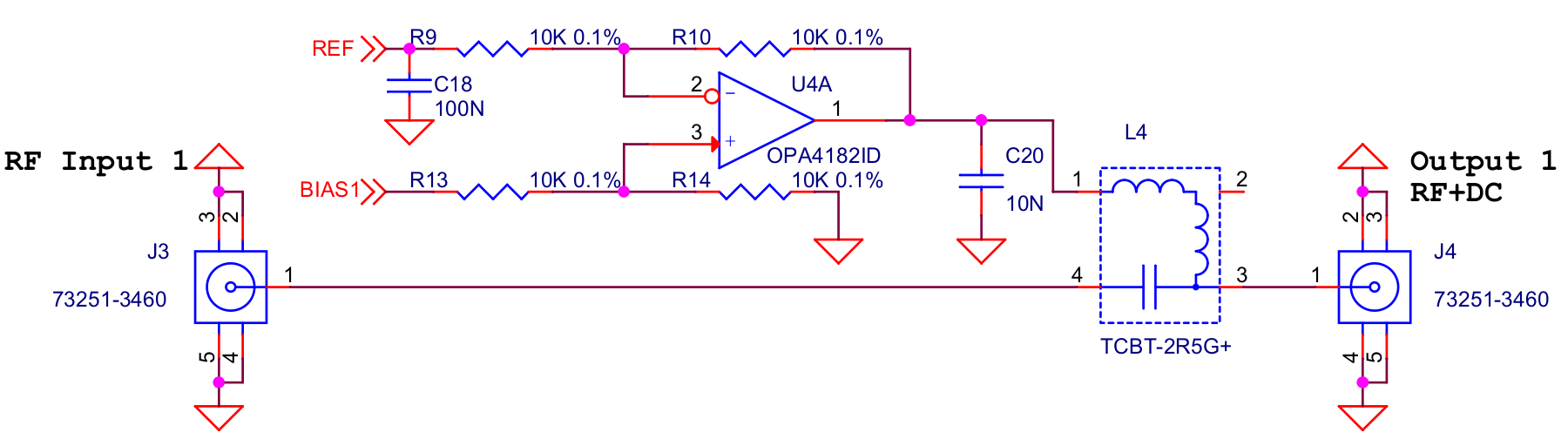}
    \caption{Schematic for the Bias Tee for combining DC and RF signals.}
    \label{fig:DC_flux_2}
\end{figure}

The RF component is synthesized by the high-speed DACs integrated within the ZCU216 boards. These RF signals are combined with the DC component using a Mini-Circuits TCBT-2R5G+ bias-tee, see Fig. \ref{fig:DC_flux_2}, yielding a single coherent output line suitable for qubit manipulation.

\subsubsection{Custom Synchronization Signaling Bus}

\begin{figure}[!ht]
    \centering
    \includegraphics[width=0.5\linewidth]{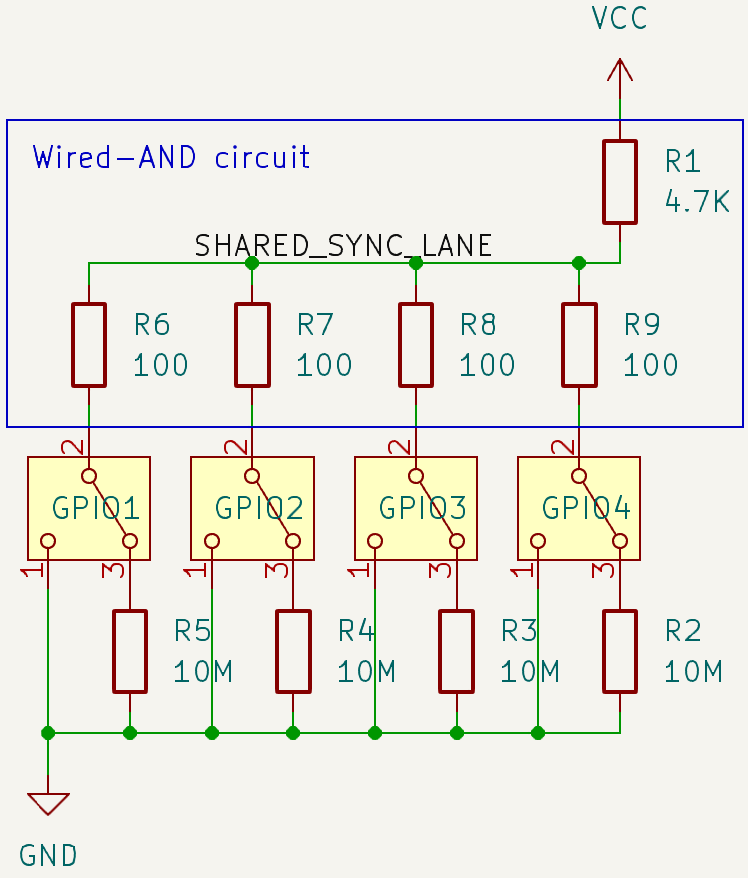}
    \caption{Schematic of Wired-AND circuit to detect when all boards are simultaneously in their ready state.}
    \label{fig:wired_AND_2}
\end{figure}

A simple external circuit was developed, together with a Sync IP, to support synchronization of programs executed by tProcessors across multiple RFSoC boards. The circuit implements a wired-AND circuit by monitoring a shared GPIO line connected to all boards, as in Fig. \ref{fig:wired_AND_2}. When each board enters the 'ready to continue' state, it signals it by driving its GPIO pin to high-impedance mode, only when all boards are in this state does the shared line in the board reach a logic high level and, therefore, informs each board that all the others are also ready to continue. This synchronization scheme is scalable and requires only one GPIO wire per board connected to this circuit.

\subsection{Firmware}

\subsubsection{Clock Tree}

Complex FPGA designs incorporate a wide range of functional blocks that operate at different clock frequencies, requiring a well-structured clocking subsystem to maintain precise coordination. The clock tree serves this role by generating and distributing synchronized clock signals across all components, ensuring deterministic timing and alignment.

\begin{figure}
    \centering
    \includegraphics[width=\linewidth]{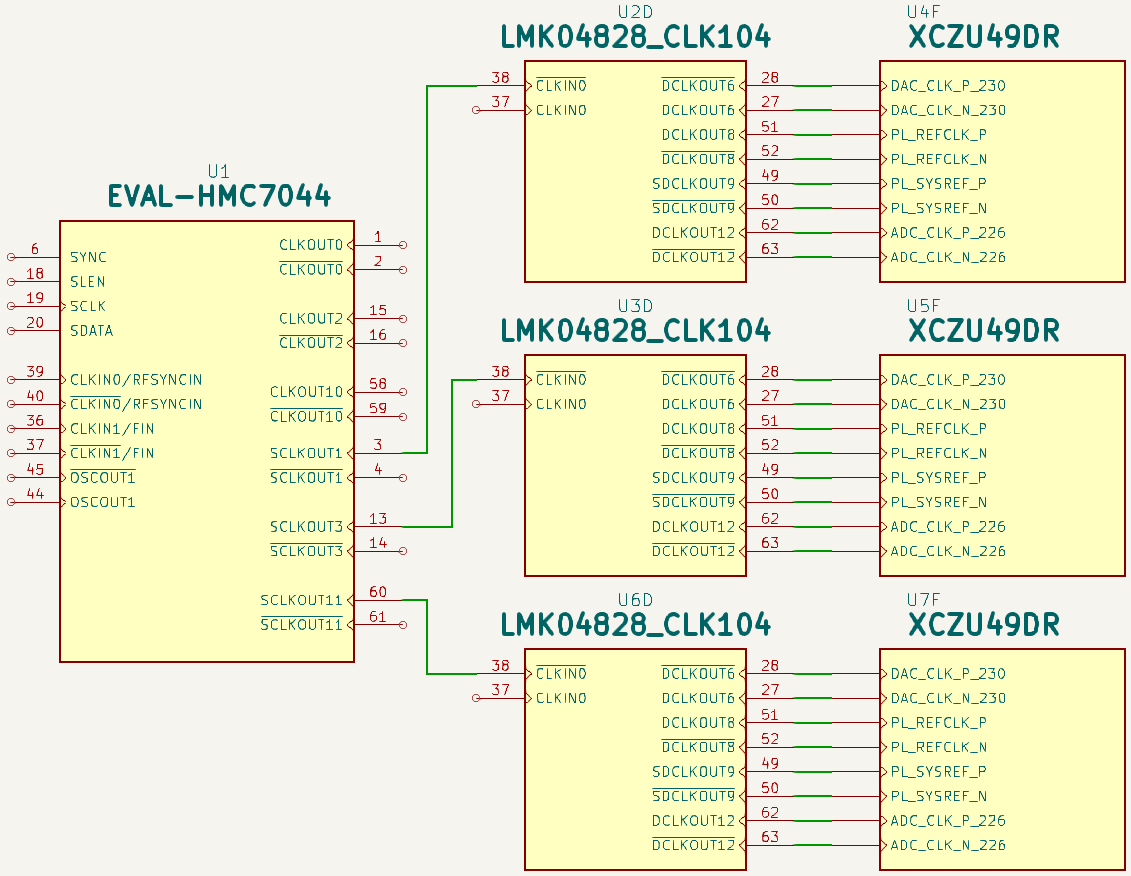}
    \caption{Clock Tree Design Part 1 (Clock generation and distribution)}
    \label{fig:clock_tree_1}
\end{figure}

In our design, the HMC7044 clock generator produces phase-coherent 7.68 MHz reference clocks, which are delivered to all boards via length-matched coaxial cables. These signals feed into the CLKin0 input of each LMK04828 chip on the CLK104 distribution boards, Fig. \ref{fig:clock_tree_1}.

Each LMK04828 then generates four synchronized clock outputs, DCLKout6, DCLKout12, DCLKout8, and SCLKout9—which are routed to their corresponding FPGA interfaces.

\begin{itemize}
    \item \textbf{DCLKout6} (DAC\_REF), \textbf{DCLKout12} (ADC\_REF), and \textbf{SCLKout9} (SYSREF) provide reference and synchronization signals for the high-speed data converters.
    \item \textbf{DCLKout8} is used for the internal FPGA logic. After passing through an MMCM (Mixed-Mode Clock Manager), it drives multiple logic clock domains within the programmable fabric:
\end{itemize}

\begin{figure}
    \centering
    \includegraphics[width=\linewidth]{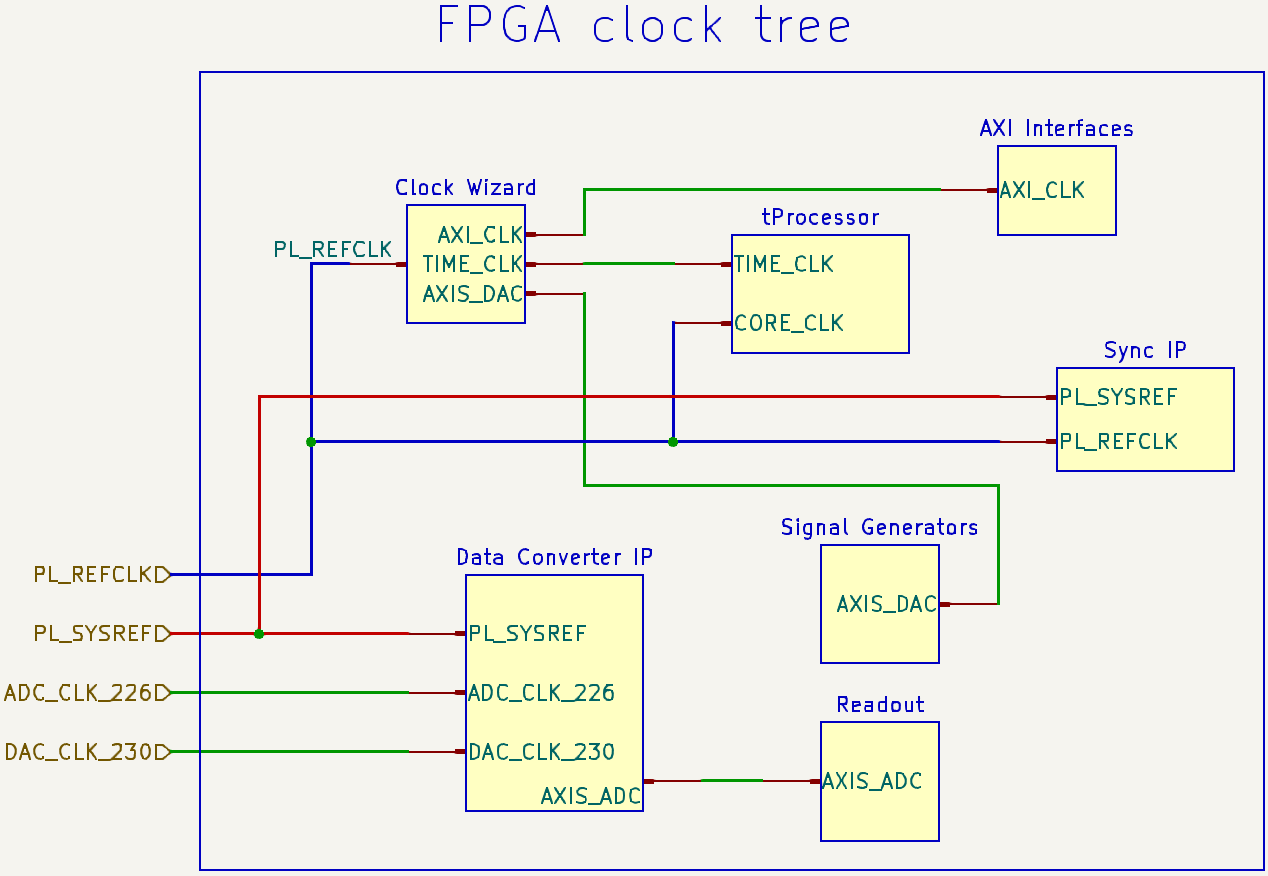}
    \caption{Clock Tree Design Part 2 (Clock generation within each FPGA)}
    \label{fig:clock_tree_2}
\end{figure}

DCLKout8 is dedicated to internal FPGA logic and, though a MMCM (Mixed-Mode Clock Manager), it serves multiple roles within the programmable logic, as seen in Fig. \ref{fig:clock_tree_2}:

\begin{enumerate}
    \item \textbf{pl\_refclk} (122.88 MHz): Clocking the MMCM, tProcessor and Sync IP.
    \item \textbf{axi\_aclk} (122.88 MHz): Clocking AXI interfaces between the processing system and programmable logic, as well as intra-logic communications.
    \item \textbf{axis\_aclk} (368.64 MHz): Driving AXI stream data interfaces between the FPGA and high-speed DACs.
    \item \textbf{time\_clock} (368.64 MHz): Controlling the tProcessor dispatcher responsible for precise waveform sequencing.
\end{enumerate}

This hierarchical clocking scheme supports timing alignment throughout the system and enables synchronous operation of up to six RFSoC boards from a single HMC7044 evaluation board. Additional distribution hardware can be used to scale beyond this configuration.

\subsubsection{Multi-Tile Synchronization}

The ZCU216 RFSoC boards consist of multiple data converters organized into tiles, each containing groups of four DACs or ADCs. Each tile shares a common clocking and data infrastructure, resulting in fixed sampling rates and latency within a tile. For applications requiring multiple tiles or multiple boards that operate in unison, achieving precise alignment of the converter latency is essential. To address this requirement, we utilize the Multi-Tile Synchronization (MTS) feature provided by AMD’s Zynq UltraScale+ RFSoC devices. This feature leverages a dedicated SYSREF signal to ensure deterministic and repeatable alignment across tiles and devices. Some steps in our implementation include:

\begin{enumerate}
    \item Distributing a single analog SYSREF input internally to all RF-ADCs and RF-DACs.
    \item Utilizing the synchronization state machine integrated within the data converter IP, activated through the RF Data Converter (RFDC) driver API.
    \item Modifying the original connection scheme between the signal generators and the Data Converter IP to meet the requirements of the MTS configuration.
\end{enumerate}

QICK standard project does not use MTS, so it was implemented in our design between all drive and readout tiles since they all work in IQ mode and share the same sampling rate (5.89824 GSPS).

\subsubsection{Sync IP}

Once the clocks of all boards are synchronized, the next challenge is to ensure that program execution across all tProcessors begins simultaneously and remains synchronized throughout the experiment.

Quantum experiments and algorithms typically involve repeated execution of pulse sequences to build up statistics. Since each board is responsible for generating sets of pulses for its assigned qubits, the programs running on each tProcessor may differ in structure and duration. 

For simple sequences, one could pad programs with wait instructions to equalize runtime. However, this approach fails when the program flow includes conditional logic, which introduces runtime variability.

To address this, we implemented a mechanism for dynamic synchronization during program execution. A new \textit{sync} instruction was introduced into the tProcessor's instruction set to support flexible, programmatic resynchronization at arbitrary points in the pulse sequence. This instruction performs two operations: it signals readiness by driving a digital output line and then waits for a system-wide signal indicating that all other tProcessors have also reached their synchronization point.

The tProcessor in the original QICK design includes multiple trigger outputs and an external flag input. We leverage these existing signals to implement the \textit{sync} instruction: upon reaching this instruction, the tProcessor asserts a dedicated trigger output to signal readiness and then stalls until its external flag input is activated. This flag input of the tProcessor is driven by a custom IP block (Sync IP) implemented in the programmable logic of each board to guarantee synchronization.

Each board's Sync IP monitors the status of the others using a single GPIO pin configured as input/output, connected to the Synchronization Signaling Bus described in the hardware section. As depicted in Fig. \ref{fig:wired_AND_2}, a board signals readiness by setting its GPIO pin to a high-impedance state, allowing the shared line to be pulled high externally. The line reaches a logic high level only when all boards are in this state. This condition is detected by all boards and is used to simultaneously release the tProcessors from their waiting state. The FSM within each Sync IP generates this release signal with deterministic timing, synchronized to the system reference clock. The time counters of the tProcessor are also reset to preserve deterministic alignment during subsequent execution.

This program synchronization mechanism can take place once at the beginning of the program or at the start of every experimental repetition. Periodic resynchronization is particularly important when program paths include conditional branches or when repetition timing varies due to measurement-based feedback.

This approach combines a dedicated instruction, a lightweight synchronization IP, and a simple hardware interconnect—enables scalable, fine-grained synchronization of distributed tProcessors and ensures coherent operation of multi-qubit experiments across multiple FPGA boards.

\subsubsection{tProcessor Modifications}

Achieving synchronization at the sub-100 picosecond level requires not only aligned clock signals and simultaneous arrival of external flags to each tProcessor, but also the assurance that all tProcessors read the external flag signal at the exact same clock cycle.

The standard version of the tProcessor does not guarantee simultaneous evaluation of external flags across boards due to the internal behavior of its instruction fetch mechanism. Specifically, the tProcessor CPU employs two control signals, 'flush' (to control the reading of the next instruction from memory) and 'fetch\_en' (to control the updating of the program counter (PC)). Since our custom synchronization instruction is conditional, when evaluated, the CPU introduces a mandatory two-cycle flush (do not execute any instruction) to prevent data conflicts resulting from a potential jump in the PC.

While this is functionally correct, in our particular case, this results in a timing uncertainty of +/- 2 clock cycles (+/- 2/122.88MHz = +/- 16 nanoseconds), depending on the lenght of each tProcessor's program, which is too large given our synchronization requirements.

To eliminate this variability, two firmware-level modifications were introduced: 1) the PC updating is paused by setting 'fetch\_en = 0' immediately after the Sync() instruction is read from memory, and 2) execution remains paused until the external flag signal from the Sync IP is received ('flag\_i == 1'b1' -> 'fetch\_en = 1'). This ensures that all tProcessors resume program execution on the same clock edge, achieving deterministic alignment.

With this final modification, we guarantee that the following conditions are met:
\begin{enumerate}
    \item The core clocks of all tProcessors are fully synchronized (via the HMC7044 and LMK04828 configurations).
    \item The external flag signals reach all tProcessors simultaneously (enabled by the Sync IP and Sync instruction).
    \item The external flags are read exactly on the next clock that they arrive across all tProcessors.
\end{enumerate}

\subsubsection{Other firmware customizations}

The QPU characterized in this work consists of 10 flux-tunable qubits arranged in a ladder configuration with fixed couplings. This architecture requires 10 drive lines, 10 flux lines and 2 readout lines. Qubits frequencies span 5-6GHz, while resonator frequencies are centered around 7.5GHz. 

To support the precise waveform shaping needed for flux pulses used in two-qubit gate implementations, we dedicated full-speed signal generators to all flux lines. For drive lines, interpolated signal generators were used to reduce FPGA resource usage.

Each board controls 5 qubits. One multiplexed signal generator and one multiplexed readout block per board were sufficient for control and measurement.

All signal generators operate at a sample rate of 5.89824 GSPS. At this rate, qubit drive tones fall into the second Nyquist zone and resonator tones into the third. Both are located more than 1 GHz away from their respective Nyquist zone edges to minimize spectral distortion.

On the acquisition side, the resonator signals are aliased from the seventh Nyquist zone into the 50–300 MHz band.

Firmware development started from the ‘qick\_tprocv2\_216\_standard’ firmware provided by the QICK \cite{qick_tprocv2_216_standard}. The following customizations were implemented to meet our system requirements:
\begin{enumerate}
    \item Six interpolated signal generators connected to DAC channels dedicated to drive pulses at a sample rate of 5.89824 GSPS.
    \item Six full-speed signal generators connected DAC channels specifically for RF flux pulses, for qubits (or couplers), operating at 5.89824 GSPS.
    \item One DAC channel for multiplexed readout operations, capable of simultaneously generating up to eight tones at 5.89824 GSPS.
    \item One ADC channel used for multiplexed read-in operations, sampling up to eight channels simultaneously at 2.4576 GSPS.
    \item Removal of unused IP blocks from the standard QICK design to reduce resource usage and simplify integration: DDR4, Dynamic Readouts (only multiplex RO are used) and replacing some 'Interpolated SG' by 'full-speed SG' (possible by reducing the sampling rate).
    \item Integration of SPI drivers functionality to control our DC AFE board from the PS using Python.
\end{enumerate}

\subsection{Software} 

\subsubsection{Petalinux and PYNQ}

To enable high-level access to the new hardware and firmware components developed in this work, we developed our Board Support Package (BSP) for the ZCU216 platform. A BSP defines the hardware configuration and low-level software components required to boot and operate an embedded Linux system on a specific target board. To build this environment, we used PetaLinux \cite{AMD_PetaLinux_UG1144}, AMD’s toolchain for developing and customizing embedded Linux distributions for Zynq and RFSoC devices. The application running on the processing system is based on PYNQ and depends on this custom Linux image for proper hardware access.

As part of the BSP development, we modify the standard device tree to support key system functionalities:

\begin{enumerate}
	\item SPI-based control of the custom DC flux board directly from the Processing System (PS/EMIO).
	\item I2C-based read/write access to the LMK04828 clock conditioner registers.
	\item GPIO Support to allow SPI register readback from LMK04828 to user space (PS/EMIO).
	\item Support for booting the system from an SD card with the custom PYNQ image.
\end{enumerate}

After validating the custom Linux image built with the modified BSP, we installed the RFSoC-PYNQ framework. To complete the system configuration, we enabled Multi-Tile Synchronization (MTS) support from within the PYNQ environment by extending the functionality of the \textit{xrfdc} \cite{xilinx_pynq_xrfdc} Python library using the RFSoC-MTS repository \cite{Xilinx_RFSoC_MTS}.

\subsubsection{Integration with Qibo}
To provide a unified software environment for quantum algorithm development and experiment execution, we integrated our control electronics with Qibo \cite{efthymiou2021qibo}, TII’s open-source framework for quantum software and hardware. Qibo is composed of a suite of interoperable libraries spanning the quantum software stack. At its core, the Qibo library supports the construction and simulation of quantum circuits, while QiboLab \cite{efthymiou2024qibolab} provides a transpilation layer that converts high-level gates into calibrated pulse sequences. It also includes hardware drivers for pulse generation and scheduling.

Complementing these layers is QiboCal \cite{pasquale2024qibocal}, a dedicated library for the characterization and benchmarking of quantum devices. It includes standardized routines for calibrating single- and two-qubit gates, as well as tools for readout fidelity assessment and coherence measurements.

To enable seamless use of our control hardware with this stack, we implemented a hardware driver that integrates directly with the Qibo interface. This driver incorporates the orchestration layer described in the next section, enabling synchronized control across multiple RFSoC boards. Through this integration, users can execute both quantum algorithms and characterization experiments using Qibo’s modular and extensible interface.

The driver supports multiplexed readout, hardware modulation/demodulation, arbitrary waveform playback, and precise flux-pulse shaping at rates up to 5.9 GSPS. It enables real-time sweeps of drive and flux pulse parameters including frequency, amplitude, phase, start time, and duration. Notably, it also supports high-resolution duration sweeps with step sizes as small as 0.17 ns, made possible through waveform padding implemented at the driver level.

\subsubsection{Multi-Board Orchestration}
Execution of quantum algorithms and characterization routines often requires coordinated pulse generation and measurement across multiple boards. These sequences are repeated thousands of times to accumulate statistical confidence in the measured quantum state.

To realize such experimental protocols, users must first configure the control electronics by initializing hardware parameters, uploading waveform data, and compiling a program—whether in assembly or a higher-level language—that specifies the temporal structure of control pulses and acquisition events. This program must also account for dynamic updates to experimental parameters across repetitions.

When multiple control boards are employed, their independent operation requires a mechanism to synchronize execution and manage data consolidation. While each board can run an extended version of the QICK (Quantum Instrumentation Control Kit) API independently, execution of multi-qubit circuits or characterization sequences that span across boards requires centralized orchestration. Without this, users would need to manually coordinate initialization, triggering, and data collection for each board.

To address this, we developed a multi-board orchestration layer that abstracts away the complexity of distributed control. This layer is capable of configuring each board, ensuring synchronized execution, and aggregating the results as though they were generated by a single monolithic device. The orchestration functionality is integrated into the driver for Qibo, allowing users to deploy complex multi-qubit circuits across multiple boards with minimal overhead.

\subsubsection{Performance Optimizations}
To maximize the capabilities of the RFSoC-based control system, we implemented several software-level optimizations aimed at improving both performance and scalability. 

A key improvement is the offloading of computational tasks—such as pulse-sequence compilation and real-time parameter updates—to the embedded ARM processors within the RFSoC. This approach reduces latency by minimizing host-device communication overhead and enables on-device processing of time-critical operations.

Our driver architecture also implements pipelined execution, allowing pulse-sequence processing and hardware execution to proceed in parallel. This overlapping of computation and communication significantly reduces total experiment runtime, particularly when executing large batches of parameter sweeps or iterative calibration routines.

Multithreading accelerates the compilation of pulse sequences into board-specific programs, waveforms, and configuration data. By distributing this preparation across multiple CPU cores, we reduce compile time and minimize the overall duration of experiments that cannot be executed in real time and must be iterated from the host computer.

Combined with features like sequence unrolling, parallel sweeps, and binned acquisitions, these enhancements allow for efficient execution of complex experiments across multiple synchronized boards.

\section{System Validation and Experimental Results} \label{sec:others}

\subsection{Multi-Board Clock Synchronization and Timing Alignment}

\begin{figure*}
    \centering
    \includegraphics[width=0.8\linewidth]{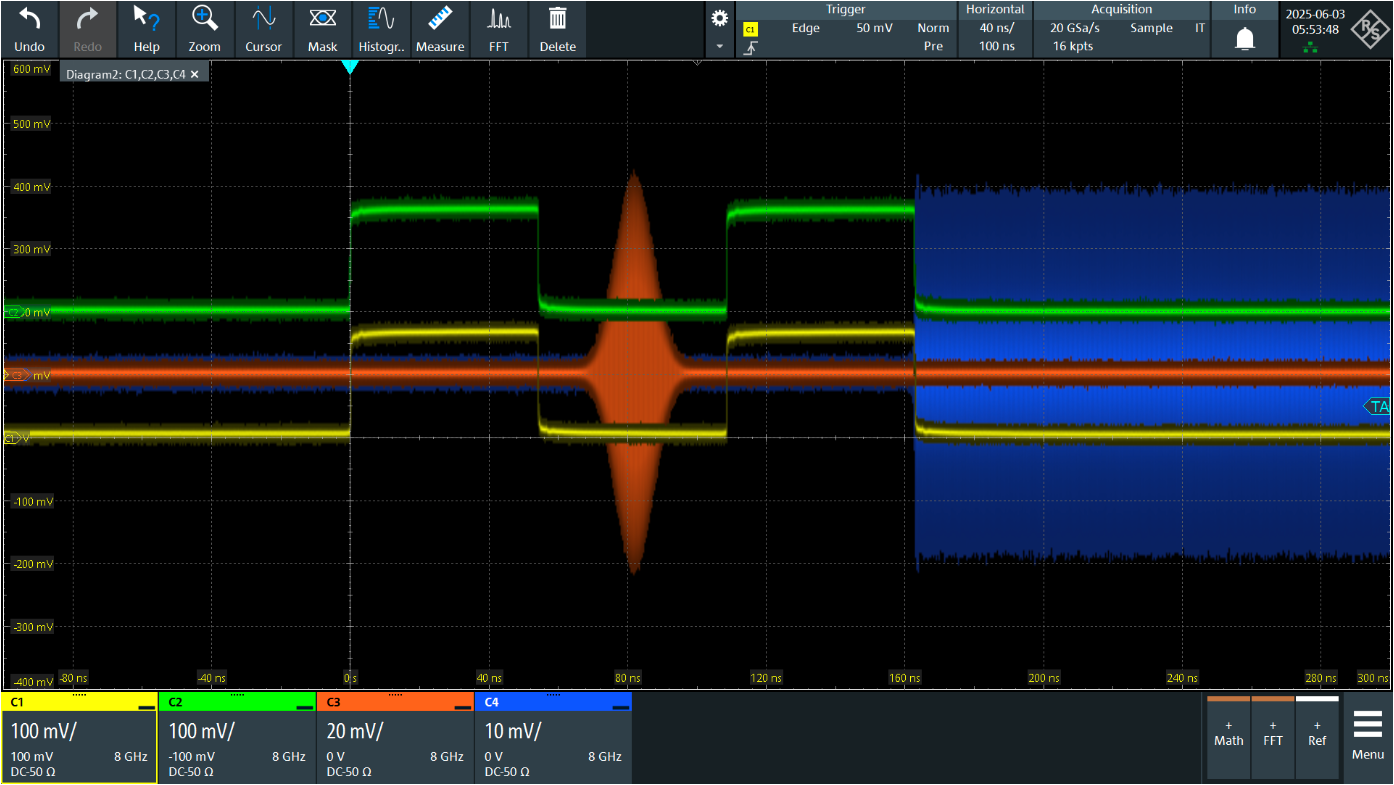}
    \caption{Synchronized flux, drive, and readout pulses across boards illustrate coordination for qubit control.}
    \label{fig:experiment}
\end{figure*}

To validate the performance of our synchronization architecture, a series of oscilloscope measurements were conducted using two ZCU216 boards, referred to as Board A and Board B, before proceeding with the qubit experiments. The following results illustrate the temporal precision and deterministic behavior of our multi-board synchronization framework under representative experimental conditions.

\begin{figure}[!ht]
    \centering
    \includegraphics[width=\linewidth]{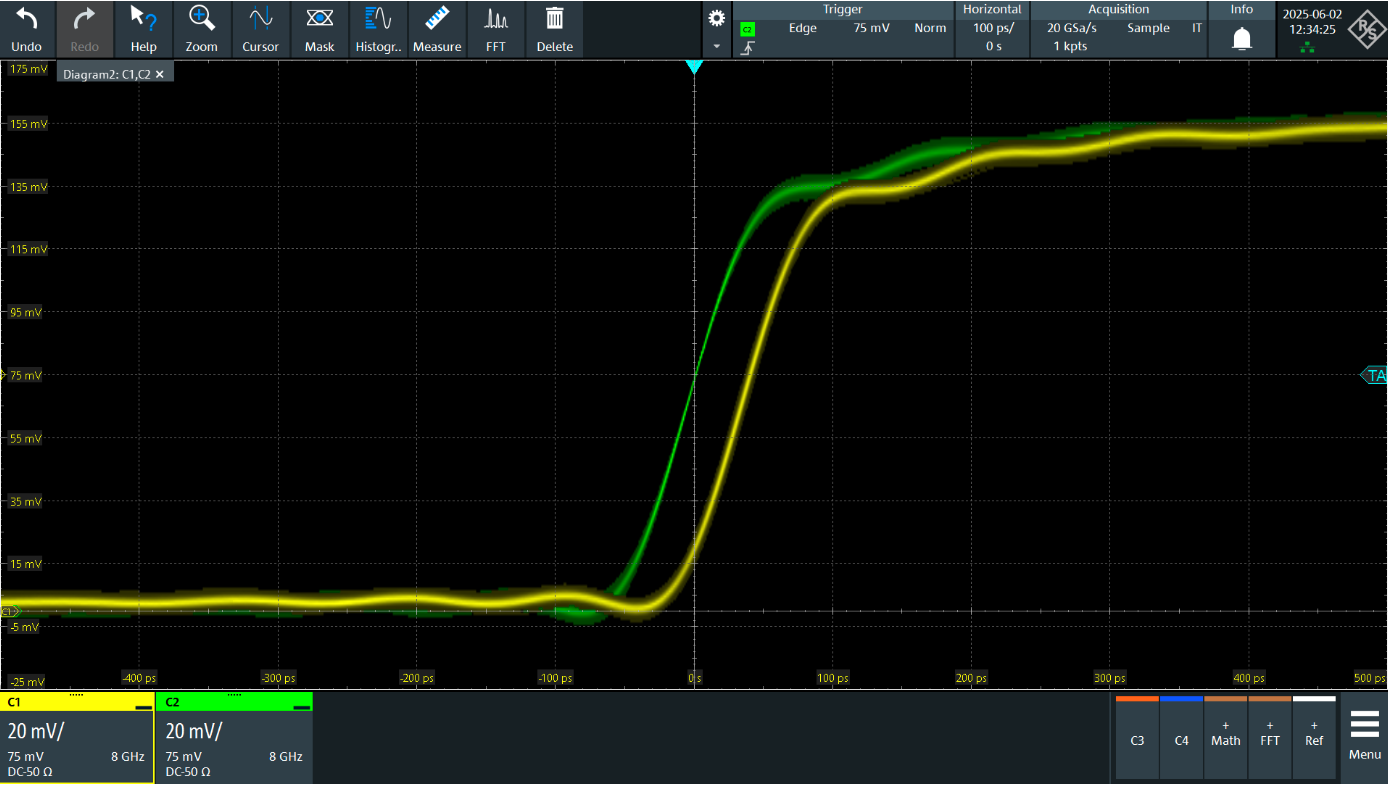}
    \caption{Overlay of square pulses from two different board shows <100 ps timing skew, demonstrating deterministic synchronization.}
    \label{fig:MBS}
\end{figure}

In the first measurement, both boards were configured to continuously output unmodulated square pulses, emulating the flux or coupler control signals typically applied in superconducting quantum processors. The outputs were aligned to initiate at t=0, and oscilloscope traces, Fig. \ref{fig:MBS}, revealed a temporal difference of less than 100 picoseconds (oscilloscope at 100 ps/div). The overlay of thousands of repetitions shows only two distinct traces, one per board, demonstrating determinism, repeatability and synchronization.

\begin{figure}[!ht]
    \centering
    \includegraphics[width=\linewidth]{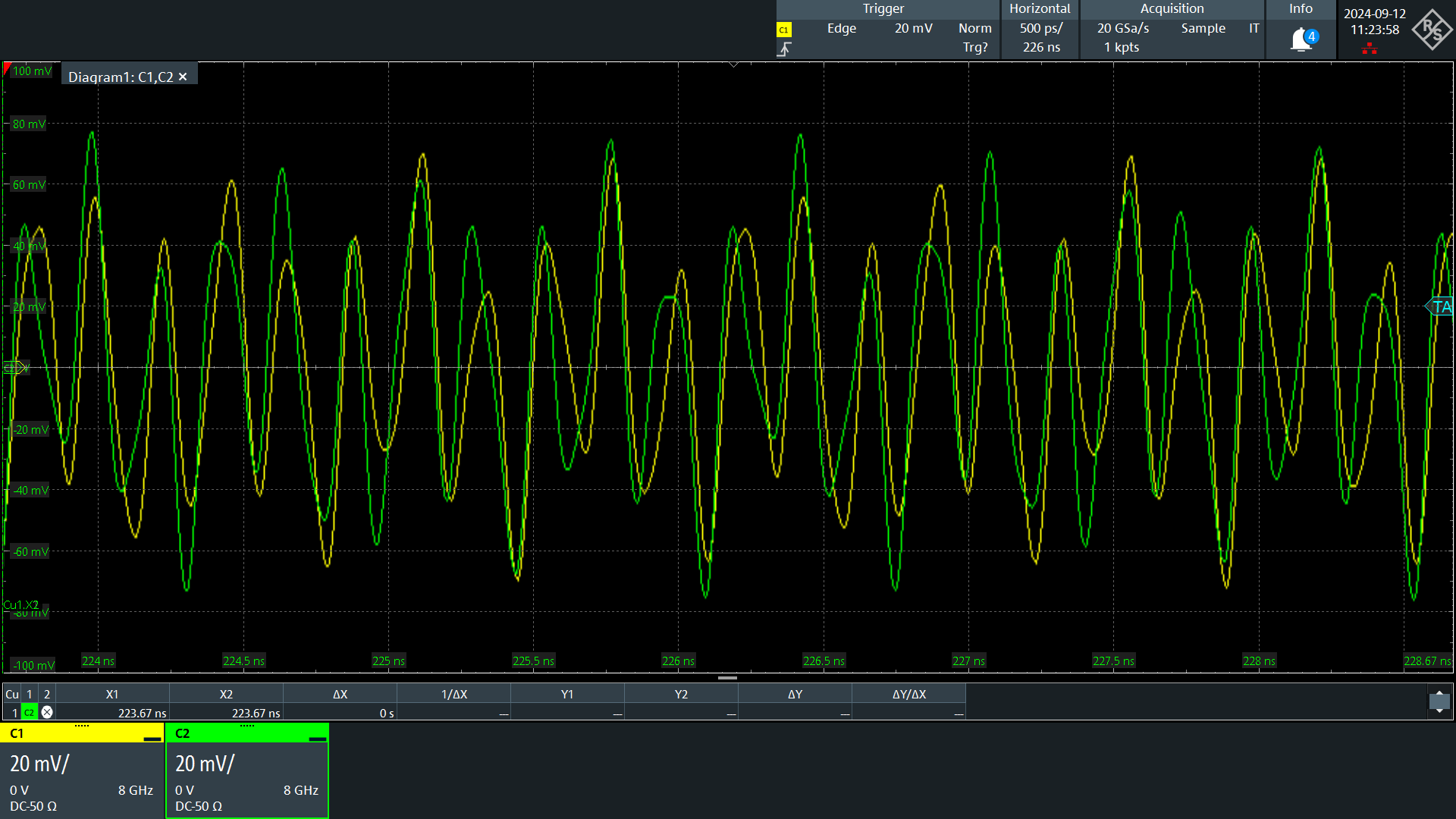}
    \caption{Phase-aligned 4.5 GHz Gaussian pulses from both boards to confirm proper Multi-Tile Synchronization.}
    \label{fig:MTS}
\end{figure}

In the second test, phase coherence was evaluated generating Gaussian-enveloped pulses modulated at 4.5 GHz, using board A and B at t=0, representative of typical qubit drive operations. Their observed phase alignment confirms that Multi-Tile Synchronization (MTS), enabled by SYSREF signal distribution, is correctly implemented across boards, as observed in Fig. \ref{fig:MTS}.

The third measurement replicates a pulse sequence, which could be an example of quantum control. The sequence includes initial flux pulses from both boards (green for board A and yellow for board B), followed by a drive pulse on board A (orange), subsequent flux pulses (green/yellow), and a final readout pulse from board B (blue). The alignment of all signals at the sub-nanosecond level, Fig. \ref{fig:experiment}, confirms the reliability of our platform for distributed control of multi-qubit experiments, where multi-board synchronization is essential for scalability.

\subsection{Experimental Results on SpinQ 10 flux-tunable qubits QPU}
As the number of qubits increases, so does the complexity of coordinating control signals across multiple hardware modules. Precise synchronization between RFSoC boards is essential for enabling coherent multi-qubit operations and scalable parallel control.

To demonstrate the capabilities of our system, we present two experiments. First, we perform a simultaneous flux-dependent resonator spectroscopy across all ten qubits, showing parallel control of all channels. While this experiment does not require precise timing alignment, it illustrates our ability to reliably operate the entire system concurrently. Second, we characterize a flux-mediated CZ-type interaction between two qubits controlled by different boards. The resulting chevron pattern serves as a sensitive probe of timing alignment, validating the temporal precision required for cross-board entangling gates.

\begin{figure*}
    \centering
    \includegraphics[width=1\linewidth]{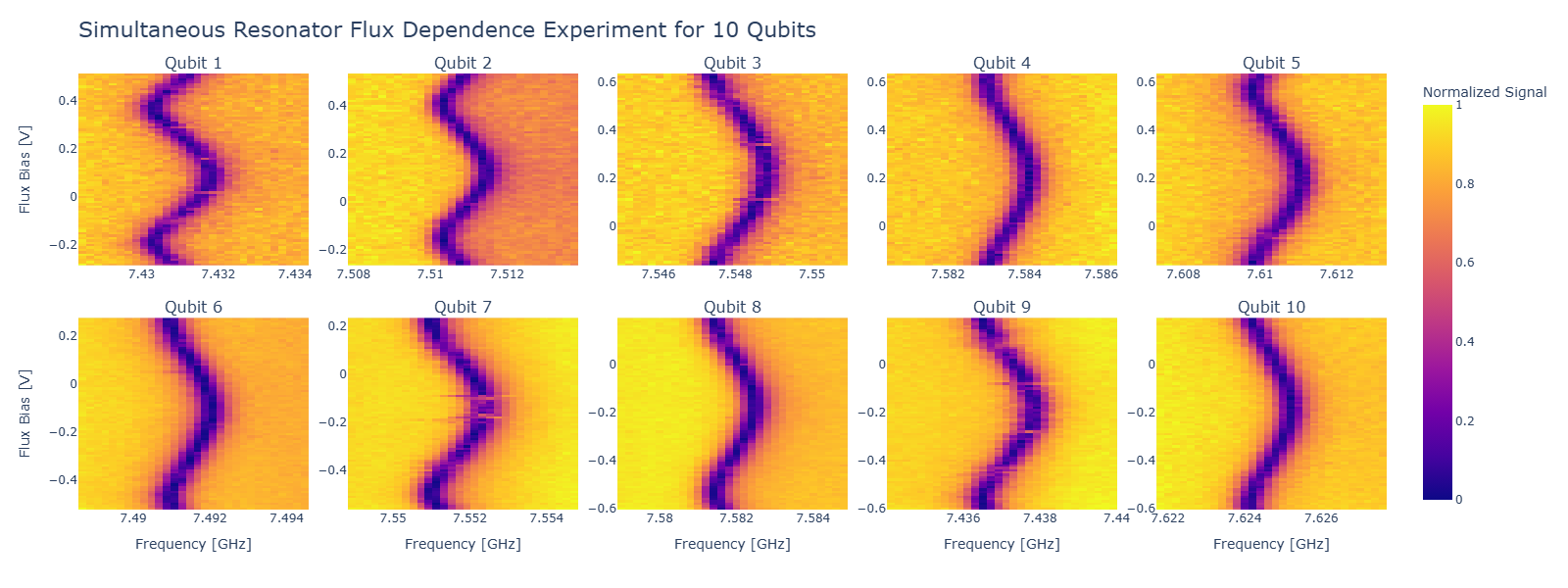}
    \caption{Results for the simultaneous execution of resonator flux dependence experiments on all 10 qubits. Each subplot corresponds to a qubit, showing a resonator response as a function of readout frequency and flux bias voltage.}
    \label{fig:flux-resonator}
\end{figure*}

\subsubsection{QPU Overview and Experimental Setup}
\begin{figure}[!ht]
    \centering
    \includegraphics[width=\linewidth]{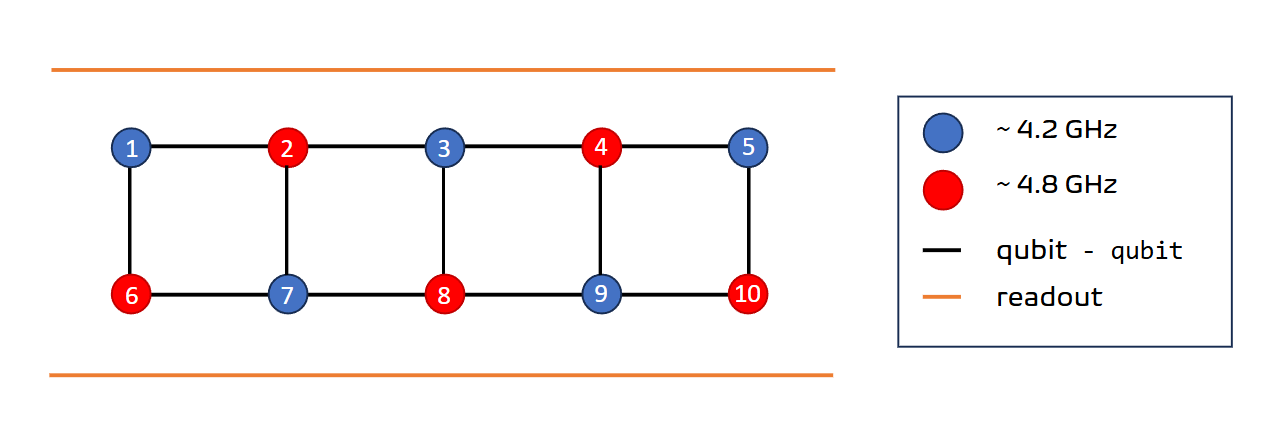}
    \caption{Layout of the 10-qubit processor used in this work. Qubits are arranged in a ladder topology with fixed capacitive couplings (black lines). Qubits alternate in frequency between ~4.2 GHz (blue) and ~4.8 GHz (red). Two readout resonators (orange lines) are shared across the upper and lower rows of qubits, enabling frequency-multiplexed readout.}
    \label{fig:spinq10q}
\end{figure}

The quantum processor used in these experiments is a 10-Qubit Shaowei Superconducting Quantum Processor purchased in 2023 from SpinQ. The chip consists of 10 flux-tunable transmon qubits arranged in a ladder topology, as shown in Fig. \ref{fig:spinq10q}. Fixed capacitive couplings connect neighboring qubits along and across the ladder. Each qubit is individually addressable through dedicated drive and flux-control lines, while readout is multiplexed across two common resonator feedlines. Qubits are staggered in frequency, alternating between approximately 4.2 GHz and 4.8 GHz.

In our current setup, control of the 10 qubits is distributed across two synchronized RFSoC boards, each handling the drive lines, flux lines, and multiplexed readout for a subset of five qubits.  This configuration enables coordinated control sequences across all qubits and forms the basis for the experiments described in the following sections. A detailed diagram of the signal routing and experimental setup is provided in Appendix \ref{sec:appA}.

\subsubsection{Simultaneous Resonator Flux Dependence Experiment}
To demonstrate full-system parallel control, we performed a simultaneous resonator flux-dependence experiment across all ten qubits. This experiment verifies that the system can coordinate concurrent waveform generation and readout across multiple RFSoC boards.

The experiment consists in performing resonator spectroscopies while sweeping the DC bias voltage applied to the flux line of the qubit. The range of voltage scanned is centered around the bias voltage that sets the qubits at their sweet spot, the point where the qubit transition frequency is first-order insensitive to flux noise.

Each qubit flux line has slightly different impedance and coupling to its dc-SQUID, resulting in different ranges of flux for the same voltage range. In qubits with high coupling we can observe the periodic response of the frequency of the qubit as a function of $\Phi / \Phi_{0}$.

Plots from Fig. \ref{fig:flux-resonator} show the magnitude of the response to a probe pulse (z axis) while varying the readout frequency (x axis) and the flux biassing voltage (y axis). Changes to the applied flux affect the frequency of the qubit which, being coupled to a resonator, result in shifts of the dressed frequency of the resonator. The frequency of the resonator is maximum when the frequency of the qubit is also maximum. At that point the qubit is first order insensitive to flux noise. Since this behavior is periodic in flux, we choose the biassing point that requires the least amount of current to reduce thermal load applied to the chip.

\begin{figure*}[!ht]
    \centering
    \subfloat[]{\includegraphics[width=\linewidth]{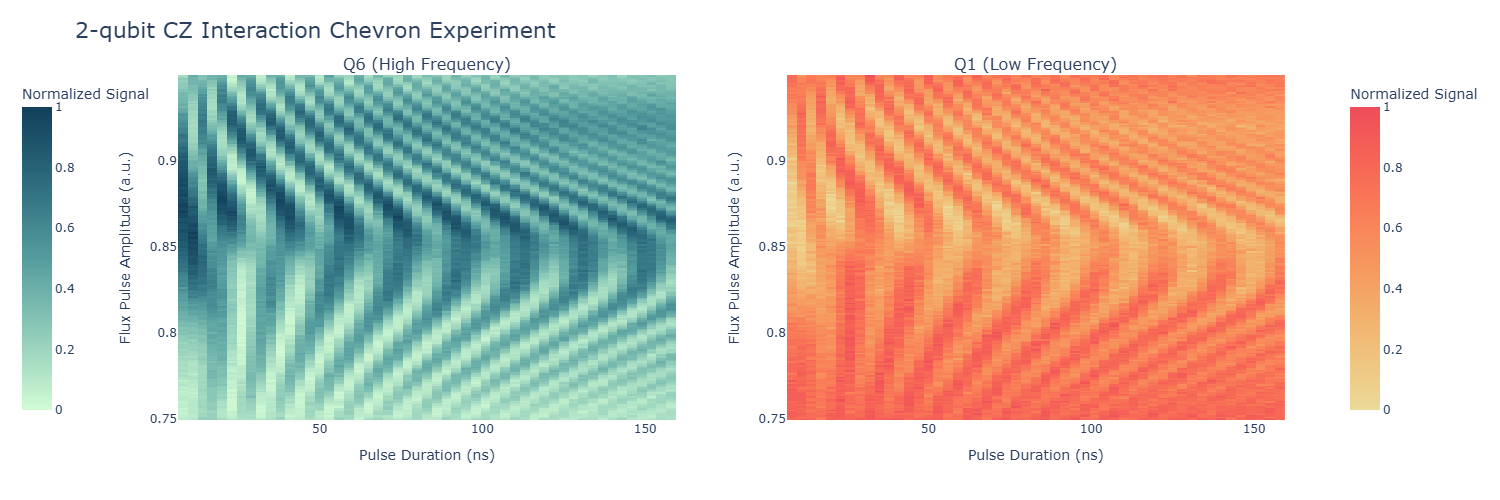}\label{fig:q1q6_CZ_a}} \\ 
    \subfloat[]{\includegraphics[width=\linewidth]{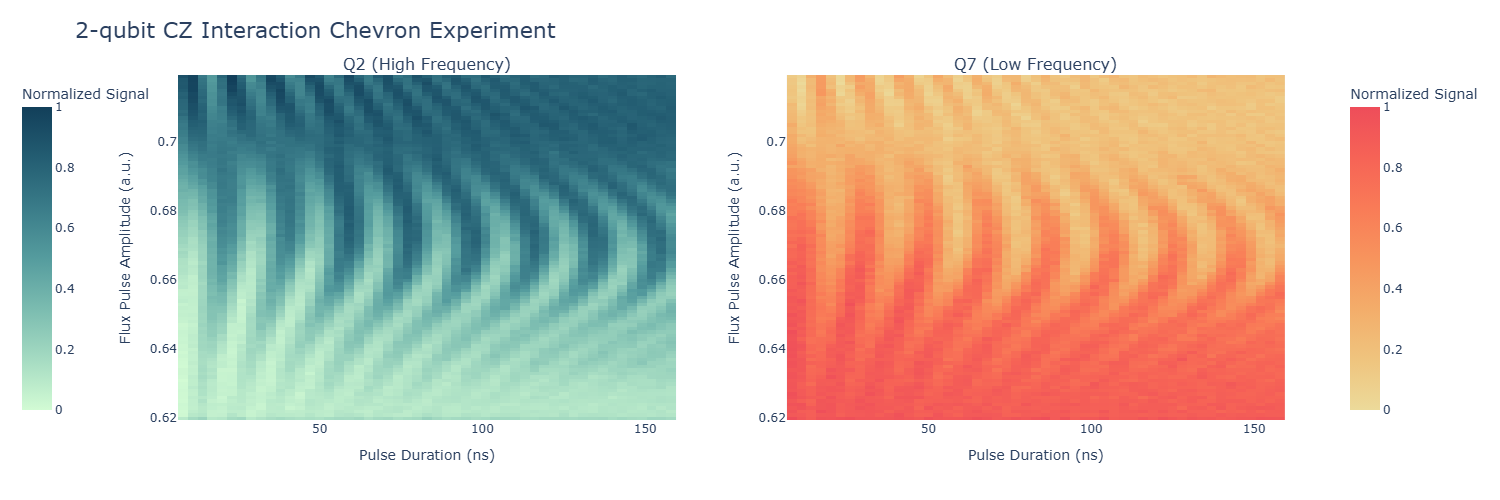}\label{fig:q2q7_CZ_b}}
    \caption{CZ interaction chevrons for pairs of qubits q2-q7 (a) and q1-q6 (b). In the z-axis we see the normalized magnitude of the readout response changing as a function of the amplitude of the flux pulse (y-axis) and the duration of the pulse (x-axis) applied to the high-frequency qubit. The low frequency qubit transitions between a state 1 (dark color) and state 0 (light color), whereas the high frequency qubit transitions between state 1 (light color) and state 2 (dark color). Each qubit in the pair is being controlled by a different board.}
    \label{fig:combined_CZ_chevrons}
\end{figure*}

\subsubsection{2-qubit CZ Interaction Chevron Experiments Between Qubits Controlled by Different Boards}
To validate our synchronization and control capabilities across multiple RFSoC boards, we performed cross-board controlled-phase (CZ) gate experiments. This class of two-qubit entangling operations plays a central role in superconducting quantum processors and is foundational for implementing quantum error correction protocols such as the surface code. We specifically selected the CZ gate as a benchmark because it is highly sensitive to pulse timing and phase alignment, especially when implemented across physically separated control electronics. High-fidelity CZ gate operation requires precise modulation of flux bias on one of the qubits to temporarily activate an interaction with a neighboring qubit, and any sub-nanosecond desynchronization can degrade gate performance by inducing spurious population transfer or unwanted phase accumulation.

The CZ gate operates by exploiting an avoided-level crossing between the $\lvert 11 \rangle$ and $\lvert 02 \rangle$ states in a coupled transmon system. During the gate, a flux pulse detunes the higher-frequency qubit towards the low-frequency one, bringing the $\lvert 11 \rangle$ and $\lvert 02 \rangle$ states close to resonance. A full coherent oscillation through this interaction imparts the conditional phase of $\pi$ on the $\lvert 11 \rangle$ state required to implement a CZ gate \cite{krantz2019quantum}.

Figure \ref{fig:combined_CZ_chevrons} shows the result of such chevron experiments for the pairs q2–q7 and q1–q6, where each qubit in the pair is controlled by a different RFSoC board. We vary the amplitude and duration of the square flux pulse applied to the high-frequency qubit and measure the response signal to infer state populations. The heatmaps exhibit clear chevron-like interference fringes, characteristic of coherent oscillations between |11⟩ and |02⟩. These fringes confirm the ability to resolve fast oscillations and capture the full population transfer and return, validating not only coherent control but also the high-fidelity timing alignment across boards.

\section{Conclusion}
We have presented Manarat, a scalable control platform for superconducting quantum processors that builds upon open-source QICK framework to support synchronized operation across multiple RFSoC boards. Through the integration of a low-jitter clock distribution network, synchronization firmware enhancements, and custom analog front-end electronics, the system achieves sub-100 ps timing alignment and deterministic program execution across boards. These capabilities were experimentally validated on a 10-qubit flux-tunable superconducting processor, demonstrating full-system simultaneous control and coherent two-qubit gate calibration across control boundaries. Our system enables synchronized multi-board operation and is fully integrated with the Qibo software stack for quantum device characterization and algorithm execution. These results establish a practical path for scaling pulse-level control systems to support mid-scale quantum processors and provide a foundation for future extensions involving feedback-enabled architectures, data communication, and other enhancements to support error-corrected quantum protocols.

\section*{Acknowledgments}
We thank the QICK development team at Fermilab for their foundational contributions. This work would not have been possible without their open-source framework, which served as the basis for our extended control system. We are especially grateful to Dr. Gustavo Cancelo and Dr. Sara Sussman at Fermilab for their continuous support and guidance on the use of QICK. Dr. Sussman has also provided invaluable advice on quantum processor characterization, experimental setup optimization, and broader scientific matters.

We gratefully acknowledge the guidance and support of Prof. Dr Frederico Brito, director of the Quantum Computing Hardware Laboratory at the Quantum Research Center, Technology Innovation Institute (QRC, TII), who played a key role in enabling this work. In addition to overseeing the lab’s overall research program and coordinating efforts across multiple groups, he provided valuable feedback through several detailed reviews of this manuscript.

We also thank Dr. Tatiana Kazieva, Mr. Alexey Zharinov, and Mr. Varun Madhavan of QRC, TII for their essential support in fridge setup, chip installation, line testing, and ongoing lab maintenance.

Dr. Javier Serrano contributed to early strategic decisions and participated in testing the QICK platform, and Mr. Andrei Paliakevich designed the TII custom analog front-end used in this work.

We would like to acknowledge the use of ChatGPT (OpenAI, 2025) for language editing, formatting assistance, and preliminary phrasing suggestions. The final text, technical content, and all scientific conclusions are solely the responsibility of the authors.

\appendix
\section{Experimental Setup} \label{sec:appA}
\begin{figure*}
    \centering
    \includegraphics[width=\linewidth]{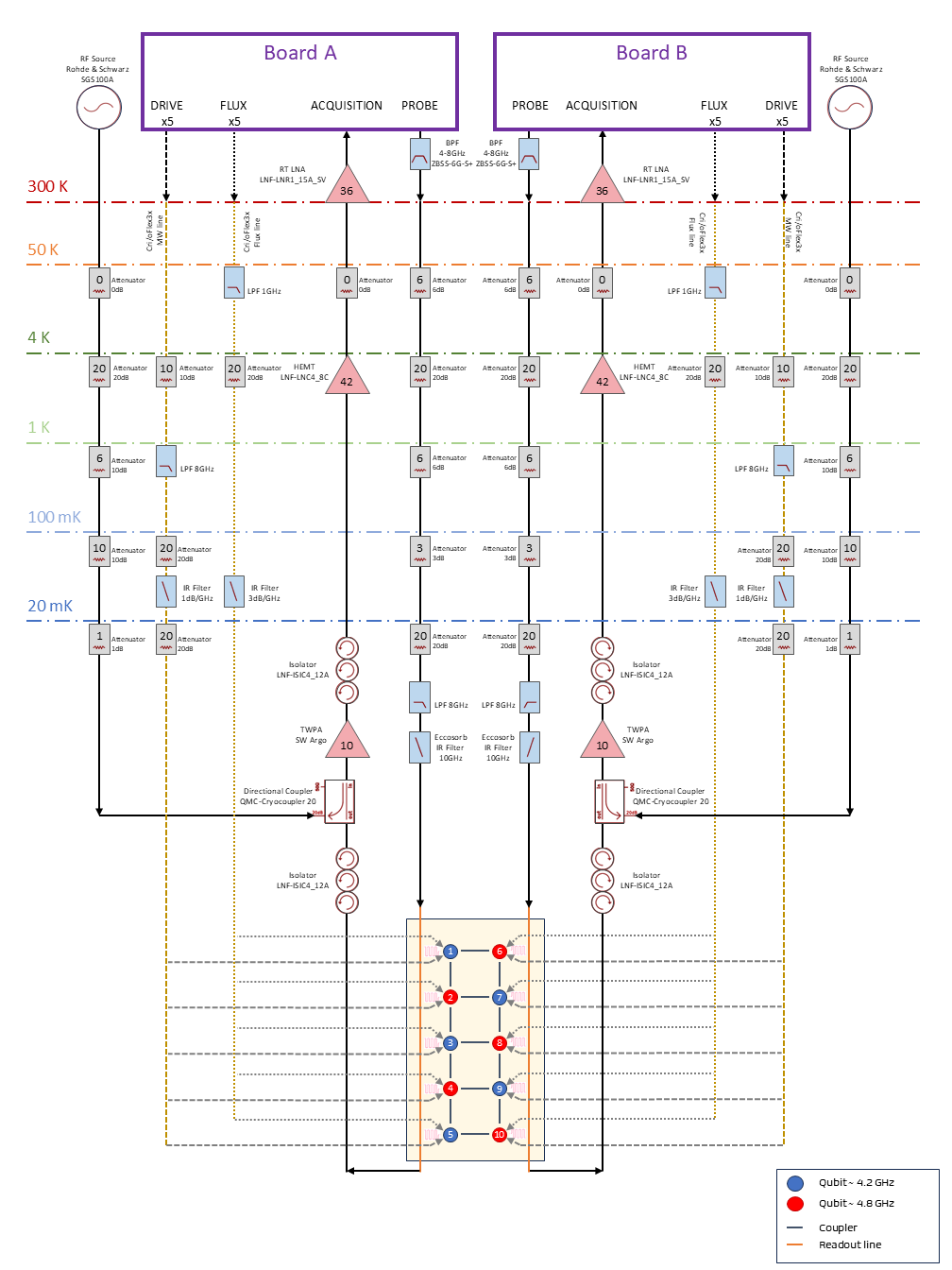}
    \caption{Full experimental setup for the 10-qubit processor.}
    \label{fig:Connectivity}
\end{figure*}

Full experimental setup for the 10 flux-tunable qubit processor can be seen in Fig. \ref{fig:Connectivity}. The diagram shows the signal routing from the two RFSoC control boards (Board A and Board B) through the room-temperature electronics, cryogenic attenuation and filtering stages, and up to the chip inside the dilution refrigerator. Each board provides five drive and five flux-control channels, along with readout and probe lines connected through traveling wave parametric amplifiers (TWPAs), isolators and low-noise amplifiers. The chip layout matches the schematic shown in Fig \ref{fig:spinq10q}, with qubits alternating in frequency and connected via fixed capacitive couplings.

\onecolumngrid
\newpage
\twocolumngrid

\section*{Bibliography}
\bibliographystyle{unsrt}  
\bibliography{main} 

\begin{thebibliography}{10}

\bibitem{shor1999polynomial}
Peter~W Shor.
\newblock Polynomial-time algorithms for prime factorization and discrete logarithms on a quantum computer.
\newblock {\em SIAM review}, 41(2):303--332, 1999.

\bibitem{arute2019quantum}
Frank Arute, Kunal Arya, Ryan Babbush, Dave Bacon, Joseph~C Bardin, Rami Barends, Rupak Biswas, Sergio Boixo, Fernando~GSL Brandao, David~A Buell, et~al.
\newblock Quantum supremacy using a programmable superconducting processor.
\newblock {\em Nature}, 574(7779):505--510, 2019.

\bibitem{preskill2018quantum}
John Preskill.
\newblock Quantum computing in the nisq era and beyond.
\newblock {\em Quantum}, 2:79, 2018.

\bibitem{kjaergaard2020superconducting}
Morten Kjaergaard, Mollie~E Schwartz, Jochen Braum{\"u}ller, Philip Krantz, Joel I-J Wang, Simon Gustavsson, and William~D Oliver.
\newblock Superconducting qubits: Current state of play.
\newblock {\em Annual Review of Condensed Matter Physics}, 11(1):369--395, 2020.

\bibitem{devoret2013superconducting}
Michel~H Devoret and Robert~J Schoelkopf.
\newblock Superconducting circuits for quantum information: an outlook.
\newblock {\em Science}, 339(6124):1169--1174, 2013.

\bibitem{gambetta2017building}
Jay~M Gambetta, Jerry~M Chow, and Matthias Steffen.
\newblock Building logical qubits in a superconducting quantum computing system.
\newblock {\em npj quantum information}, 3(1):2, 2017.

\bibitem{gottesman1997stabilizer}
Daniel Gottesman.
\newblock {\em Stabilizer codes and quantum error correction}.
\newblock California Institute of Technology, 1997.

\bibitem{fowler2012surface}
Austin~G Fowler, Matteo Mariantoni, John~M Martinis, and Andrew~N Cleland.
\newblock Surface codes: Towards practical large-scale quantum computation.
\newblock {\em Physical Review A—Atomic, Molecular, and Optical Physics}, 86(3):032324, 2012.

\bibitem{terhal2015quantum}
Barbara~M Terhal.
\newblock Quantum error correction for quantum memories.
\newblock {\em Reviews of Modern Physics}, 87(2):307--346, 2015.

\bibitem{chen2014qubit}
Yu~Chen, C~Neill, Pedram Roushan, Nelson Leung, Michael Fang, Rami Barends, Julian Kelly, Brooks Campbell, Z~Chen, Benjamin Chiaro, et~al.
\newblock Qubit architecture with high coherence and fast tunable coupling.
\newblock {\em Physical review letters}, 113(22):220502, 2014.

\bibitem{mundada2019suppression}
Pranav Mundada, Gengyan Zhang, Thomas Hazard, and Andrew Houck.
\newblock Suppression of qubit crosstalk in a tunable coupling superconducting circuit.
\newblock {\em Physical Review Applied}, 12(5):054023, 2019.

\bibitem{sung2021realization}
Youngkyu Sung, Leon Ding, Jochen Braum{\"u}ller, Antti Veps{\"a}l{\"a}inen, Bharath Kannan, Morten Kjaergaard, Ami Greene, Gabriel~O Samach, Chris McNally, David Kim, et~al.
\newblock Realization of high-fidelity cz and zz-free iswap gates with a tunable coupler.
\newblock {\em Physical Review X}, 11(2):021058, 2021.

\bibitem{versluis2017scalable}
Richard Versluis, Stefano Poletto, Nader Khammassi, Brian Tarasinski, Nadia Haider, David~J Michalak, Alessandro Bruno, Koen Bertels, and Leonardo DiCarlo.
\newblock Scalable quantum circuit and control for a superconducting surface code.
\newblock {\em Physical Review Applied}, 8(3):034021, 2017.

\bibitem{krinner2019engineering}
Sebastian Krinner, Simon Storz, Philipp Kurpiers, Paul Magnard, Johannes Heinsoo, Raphael Keller, Janis Luetolf, Christopher Eichler, and Andreas Wallraff.
\newblock Engineering cryogenic setups for 100-qubit scale superconducting circuit systems.
\newblock {\em EPJ Quantum Technology}, 6(1):2, 2019.

\bibitem{mcdermott2014accurate}
R~McDermott and MG~Vavilov.
\newblock Accurate qubit control with single flux quantum pulses.
\newblock {\em Physical Review Applied}, 2(1):014007, 2014.

\bibitem{xu2021qubic}
Yilun Xu, Gang Huang, Jan Balewski, Ravi Naik, Alexis Morvan, Bradley Mitchell, Kasra Nowrouzi, David~I Santiago, and Irfan Siddiqi.
\newblock Qubic: An open-source fpga-based control and measurement system for superconducting quantum information processors.
\newblock {\em IEEE Transactions on Quantum Engineering}, 2:1--11, 2021.

\bibitem{xu2023qubic}
Yilun Xu, Gang Huang, Neelay Fruitwala, Abhi Rajagopala, Ravi~K Naik, Kasra Nowrouzi, David~I Santiago, and Irfan Siddiqi.
\newblock Qubic 2.0: An extensible open-source qubit control system capable of mid-circuit measurement and feed-forward.
\newblock {\em arXiv preprint arXiv:2309.10333}, 2023.

\bibitem{stefanazzi2022qick}
Leandro Stefanazzi, Kenneth Treptow, Neal Wilcer, Chris Stoughton, Collin Bradford, Sho Uemura, Silvia Zorzetti, Salvatore Montella, Gustavo Cancelo, Sara Sussman, et~al.
\newblock The qick (quantum instrumentation control kit): Readout and control for qubits and detectors.
\newblock {\em Review of Scientific Instruments}, 93(4), 2022.

\bibitem{ding2024experimental}
Chunyang Ding, Martin Di~Federico, Michael Hatridge, Andrew Houck, Sebastien Leger, Jeronimo Martinez, Connie Miao, David~Schuster I, Leandro Stefanazzi, Chris Stoughton, et~al.
\newblock Experimental advances with the qick (quantum instrumentation control kit) for superconducting quantum hardware.
\newblock {\em Physical Review Research}, 6(1):013305, 2024.

\bibitem{amd_rfsoc_overview}
{AMD}.
\newblock {Zynq UltraScale+ RFSoC Overview}.
\newblock \url{https://docs.amd.com/v/u/en-US/ds889-zynq-usp-rfsoc-overview/}, 2024.
\newblock Accessed: 2025-07-14.

\bibitem{werninghaus2021leakage}
Max Werninghaus, Daniel~J Egger, Federico Roy, Shai Machnes, Frank~K Wilhelm, and Stefan Filipp.
\newblock Leakage reduction in fast superconducting qubit gates via optimal control.
\newblock {\em npj Quantum Information}, 7(1):14, 2021.

\bibitem{gustavsson2013improving}
Simon Gustavsson, Olger Zwier, Jonas Bylander, Fei Yan, Fumiki Yoshihara, Yasunobu Nakamura, Terry~P Orlando, and William~D Oliver.
\newblock Improving quantum gate fidelities by using a qubit to measure microwave pulse distortions.
\newblock {\em Physical review letters}, 110(4):040502, 2013.

\bibitem{bland20252d}
Matthew~P Bland, Faranak Bahrami, Jeronimo~GC Martinez, Paal~H Prestegaard, Basil~M Smitham, Atharv Joshi, Elizabeth Hedrick, Alex Pakpour-Tabrizi, Shashwat Kumar, Apoorv Jindal, et~al.
\newblock 2d transmons with lifetimes and coherence times exceeding 1 millisecond.
\newblock {\em arXiv preprint arXiv:2503.14798}, 2025.

\bibitem{wang2025evidence}
Qianxu Wang, Sara~Magdalena G{\'o}mez, Juan~S Salcedo-Gallo, Roy Leibovitz, Jake Freeman, Salil Bedkihal, and Mattias Fitzpatrick.
\newblock Evidence of memory effects in the dynamics of two-level system defect ensembles using broadband, cryogenic transient dielectric spectroscopy.
\newblock {\em arXiv preprint arXiv:2505.18263}, 2025.

\bibitem{amd_zcu216}
{AMD}.
\newblock {ZCU216 Evaluation Board}.
\newblock \url{https://www.amd.com/en/products/adaptive-socs-and-fpgas/evaluation-boards/zcu216.html}, 2024.
\newblock Accessed: 2025-07-14.

\bibitem{xu2025multi}
Yilun Xu, Abhi~D Rajagopala, Neelay Fruitwala, and Gang Huang.
\newblock Multi-fpga synchronization and data communication for quantum control and measurement.
\newblock {\em arXiv preprint arXiv:2506.09856}, 2025.

\bibitem{pynq_website}
{PYNQ}.
\newblock {PYNQ: Python productivity for Zynq}.
\newblock \url{https://www.pynq.io/}, 2024.
\newblock Accessed: 2025-07-14.

\bibitem{amd_xm655}
{AMD}.
\newblock {XM655 RF Analog Front-End Card}.
\newblock \url{https://docs.amd.com/r/en-US/ug1390-zcu216-eval-bd/XM655/}, 2024.
\newblock Accessed: 2025-07-14.

\bibitem{analog_devices_hmc7044}
{Analog Devices}.
\newblock {HMC7044: Jitter Attenuator and Clock Generator}.
\newblock \url{https://www.analog.com/en/products/hmc7044.html}, 2024.
\newblock Accessed: 2025-07-14.

\bibitem{ti_lmk04828}
{Texas Instruments}.
\newblock {LMK04828: Ultra-Low Jitter Clock Synthesizer and Jitter Cleaner}.
\newblock \url{https://www.ti.com/product/LMK04828/}, 2024.
\newblock Accessed: 2025-07-14.

\bibitem{xilinx_pynq_xrfclk}
{Xilinx}.
\newblock {PYNQ xrfclk GitHub Repository}.
\newblock \url{https://github.com/Xilinx/PYNQ/tree/master/sdbuild/packages/xrfclk}, 2024.
\newblock Accessed: 2025-07-14.

\bibitem{qick_tprocv2_216_standard}
QICK Developers.
\newblock Qick firmware project: qick\_tprocv2\_216\_standard.
\newblock \url{https://github.com/openquantumhardware/qick/tree/main/firmware/projects/qick_tprocv2_216_standard}, 2025.
\newblock Accessed: 2025-07-11.

\bibitem{AMD_PetaLinux_UG1144}
{AMD}.
\newblock {\em {PetaLinux Tools Documentation: Reference Guide (UG1144)}}.
\newblock {AMD}, 2025.
\newblock Accessed: 2025-07-14.

\bibitem{xilinx_pynq_xrfdc}
{Xilinx}.
\newblock {PYNQ xrfdc GitHub Repository}.
\newblock \url{https://github.com/Xilinx/PYNQ/tree/master/sdbuild/packages/xrfdc}, 2024.
\newblock Accessed: 2025-07-14.

\bibitem{Xilinx_RFSoC_MTS}
{Xilinx}.
\newblock {RFSoC-MTS: Multi-Tile Synchronization examples for Zynq UltraScale+ RFSoC}.
\newblock GitHub repository.
\newblock Accessed: 2025-07-14.

\bibitem{efthymiou2021qibo}
Stavros Efthymiou, Sergi Ramos-Calderer, Carlos Bravo-Prieto, Adri{\'a}n P{\'e}rez-Salinas, Diego Garc{\'\i}a-Mart{\'\i}n, Artur Garcia-Saez, Jos{\'e}~Ignacio Latorre, and Stefano Carrazza.
\newblock Qibo: a framework for quantum simulation with hardware acceleration.
\newblock {\em Quantum Science and Technology}, 7(1):015018, 2021.

\bibitem{efthymiou2024qibolab}
Stavros Efthymiou, Alvaro Orgaz-Fuertes, Rodolfo Carobene, Juan Cereijo, Andrea Pasquale, Sergi Ramos-Calderer, Simone Bordoni, David Fuentes-Ruiz, Alessandro Candido, Edoardo Pedicillo, et~al.
\newblock Qibolab: an open-source hybrid quantum operating system.
\newblock {\em Quantum}, 8:1247, 2024.

\bibitem{pasquale2024qibocal}
Andrea Pasquale, Edoardo Pedicillo, Juan Cereijo, Sergi Ramos-Calderer, Alessandro Candido, Gabriele Palazzo, Rodolfo Carobene, Marco Gobbo, Stavros Efthymiou, Yuanzheng~Paul Tan, et~al.
\newblock Qibocal: an open-source framework for calibration of self-hosted quantum devices.
\newblock {\em arXiv preprint arXiv:2410.00101}, 2024.

\bibitem{krantz2019quantum}
Philip Krantz, Morten Kjaergaard, Fei Yan, Terry~P Orlando, Simon Gustavsson, and William~D Oliver.
\newblock A quantum engineer's guide to superconducting qubits.
\newblock {\em Applied physics reviews}, 6(2), 2019.

\end{thebibliography}

\end{document}